\documentclass[3p,twocolumn]{elsarticle}
\pdfoutput=1

\usepackage{graphicx}
\usepackage{amsmath,amssymb,amscd}
\usepackage{multirow}
\usepackage{comment}
\usepackage{mathtools}

\usepackage{slashed}
\usepackage{tikz}

\newcommand{\be}{\begin{equation} }
\newcommand{\ee}{\end{equation}}
\newcommand{\cL}{\mathcal{L}}

\usetikzlibrary{decorations.markings }
\tikzset{
  fermionline/.style={line width=1pt,postaction={decorate},
    decoration={markings,f
      mark=at position 0.5 with {\draw[-stealth] (0,0)--(2pt,0);}}},
  bosonline/.style={line width=1pt,decorate,
    decoration={snake,amplitude=1,segment length=4}},
  higgsline/.style={line width=1pt,dashed}
}

\begin{document}

\begin{frontmatter}
%\vspace*{-5mm}
%\begin{flushright}
%MAN/HEP/2017/10
%\end{flushright}
%\vspace*{5mm}
\title{On the robustness of the primordial power spectrum \\ in renormalized Higgs inflation}

\author[ma]{Fedor Bezrukov}
\ead{Fedor.Bezrukov@manchester.ac.uk}  

\address[ma]{The University of Manchester, School of Physics and Astronomy, \\
Oxford Road, Manchester M13 9PL, United Kingdom}

\author[hd]{Martin Pauly} 
\ead{m.pauly@thphys.uni-heidelberg.de} 

\author[hd]{Javier Rubio}
\ead{j.rubio@thphys.uni-heidelberg.de}  

\address[hd]{Institut f\"ur Theoretische Physik, Ruprecht-Karls-Universit\"at Heidelberg, \\
Philosophenweg 16, 69120 Heidelberg, Germany}

\begin{abstract}
We study the cosmological consequences of higher-dimensional operators respecting the asymptotic symmetries of the tree-level Higgs inflation action. The main contribution of these operators to the renormalization group enhanced potential is localized in 
a compact field range, whose upper limit is close to the end of inflation. The spectrum of primordial fluctuations in the 
so-called  \textit{universal regime} turns out to be almost insensitive to radiative corrections and in 
excellent agreement with the present cosmological data. However, higher-dimensional operators can play
an important role in \textit{critical Higgs inflation} scenarios containing a quasi-inflection point along the 
inflationary trajectory. The interplay of radiative corrections with this quasi-inflection point may
translate into a sizable modification of the inflationary observables.
\end{abstract}


\end{frontmatter}
%\maketitle
 %%%%%%%%%%%%%%%%%%%%%%%%%%%%%
\section{Introduction}\label{sec:intro}
%%%%%%%%%%%%%%%%%%%%%%%%%%%%%

The recent results of the Planck collaboration favor the simplest
realization of the inflationary paradigm: a single-field model of inflation with a Gaussian 
spectrum of primordial perturbations
and no isocurvature perturbations~\cite{Ade:2015lrj}. The $\Lambda$CDM scenario is also remarkably
successful: a spatially flat Universe containing a proper mixture of baryons, neutrinos and cold 
dark matter can easily account for observations in a vast range of epochs and scales. On top of the
cosmological probes, collider experiments point towards 
a minimal particle physics scenario containing only a $125$-$126$ GeV scalar field with properties
closely resembling those of the Standard Model (SM) Higgs \cite{Aad:2012tfa,Chatrchyan:2012ufa}. Nothing 
beyond this minimalistic framework has been discovered so far. In the absence of new degrees of freedom, it is 
interesting to look for scenarios able to describe the observed Universe using the existing  particle 
content. 

In this paper we consider the observational consequences of Higgs inflation in the presence of  higher-dimensional operators 
compatible with the symmetries of the tree-level action. In Higgs inflation, the role of the inflaton field is played by the SM Higgs non-minimally coupled to 
gravity. No additional degrees of freedom beyond those already present in the SM are introduced to explain the
background properties of the Universe, or the generation of an almost scale-invariant spectrum of 
primordial fluctuations.

The role of scalar fields non-minimally coupled to gravity has been extensively 
explored in the literature \cite{Adler:1982ri, Minkowski:1977aj, Smolin:1979uz, Zee:1978wi, Salopek:1988qh}. 
Higgs inflation particularizes the non-minimally coupled field to the SM Higgs \cite{Bezrukov:2007ep}.
For a metric signature $(-,+,+,+)$, the relevant part of the Higgs inflation action 
reads \cite{Salopek:1988qh, Bezrukov:2007ep}
 \begin{equation}
\label{lagr}
S = \int d^4x \sqrt{-g} \left[ \frac{f(h)}{2}   R -\frac{1}{2}\left(\partial h\right)^2- U(h) \right] \,,
\end{equation} 
with $f(h) = M_P^2 + \xi h^2$, 
$h$ the Higgs field in the unitary gauge and
\begin{equation}
U(h)=\frac{\lambda}{4}(h^2-v_{\rm EW}^2)^2\,,
\end{equation}
the SM symmetry breaking potential. The non-minimal coupling $\xi$ is assumed to be in the 
range $1\ll \xi\ll M_P^2/v^2_{\rm EW}$ with $v_{\rm EW}\simeq 250$ GeV the vacuum expectation value of 
the Higgs field and $M_P=2.435 \times 10^{18}$~GeV the reduced Planck mass. 

The universal predictions of Higgs inflation are intimately related to the approximate scale-invariance 
of Eq.~\eqref{lagr} at $h\gg M_P/\sqrt{\xi}$. This emergent symmetry translates into a scale-invariant
spectrum of primordial fluctuations in excellent agreement with the latest Planck results \cite{Ade:2015lrj}.  
The embedding of Higgs inflation in a fully scale-invariant framework was considered in Refs.~\cite{GarciaBellido:2011de,Bezrukov:2012hx,Karananas:2016kyt}.

The non-minimal coupling to gravity makes Higgs inflation non-renormalizable 
\cite{Barbon:2009ya,Burgess:2009ea,Burgess:2010zq,Bezrukov:2010jz}, which in turn forbids the interpretation of the model as a 
UV-complete theory.  Thus, Higgs inflation should be
understood as an effective field theory (EFT) applicable below a given cutoff  
$\Lambda$ \cite{Bezrukov:2010jz,George:2015nza}. This scale could signal the onset of a strong coupling 
regime \cite{Aydemir:2012nz,Calmet:2013hia,Escriva:2016cwl} or the appearance of additional degrees of freedom 
\cite{Giudice:2010ka,Barbon:2015fla}. The cutoff following from the tree-level action \eqref{lagr} is not a 
fixed scale, but rather a dynamical quantity that depends on the expectation value of the Higgs field \cite{Bezrukov:2010jz} 
\begin{equation}\label{Lcut}
\Lambda(h)=\frac{M_P^2+\xi(1+6\xi)h^2}{\xi\sqrt{M_P^2+\xi h^2}}\,.
\end{equation}
As shown in Ref.~\cite{Bezrukov:2010jz}, all energy scales involved in the inflationary and  post-inflationary evolution of 
the Universe are parametrically smaller than $\Lambda(h)$. The weak-coupling approximation therefore remains valid 
(see also Refs.~\cite{Ferrara:2010in,Moss:2014nya,Ren:2014sya}). 
 
A sensible computation of radiative corrections within the EFT interpretation of Higgs inflation requires the 
introduction of an infinite number of higher-dimensional operators suppressed by the cutoff scale in Eq.~\eqref{Lcut}. 
The minimal set of operators is generated by the theory itself via radiative corrections. By construction, these 
operators respect the asymptotic scale-symmetry of the  classical action at large field values. However, there is a larger set of operators with this property. Indeed any higher-dimensional operator constructed out of the Higgs field, the cutoff scale $\Lambda(h)$ and some numerical coefficients $c_n$
\begin{equation}\label{genope}
\frac{c_n\,{\cal O}_{n}(h)}{\left[\Lambda(h)\right]^{n-4}}\,,
\end{equation}
becomes automatically scale-invariant in the large field regime $h\gg M_P/\sqrt{\xi}$.

The predictions of Higgs inflation can only be considered robust if they are insensitive to the particular 
choice of operators and coefficients in Eq.~\eqref{genope}. The sensitivity of the inflationary observables to 
higher-dimensional operators has been partially addressed in the literature. The analysis presented in 
this paper complements the results of Refs.~\cite{Fumagalli:2016lls} and \cite{Enckell:2016xse}. In particular, 
Ref.~\cite{Fumagalli:2016lls} considers Higgs-inflation scenarios away from the so-called \textit{critical case}, while 
Ref.~\cite{Enckell:2016xse} does not consider the field-dependence of radiative corrections. In this paper we extend the   
analysis of these references to different ultraviolet (UV) completions and to the \textit{critical Higgs-inflation} 
regime \cite{Bezrukov:2014bra,Hamada:2014iga} containing a \textit{quasi}-inflection point along the inflationary trajectory.

This paper is organized as follows. In Section \ref{sec:HI} we reformulate Higgs inflation in the so-called 
Einstein frame, where the gravitational sector takes the usual Einstein-Hilbert form. The renormalization 
procedure for this (intrinsically non-renormalizable)  theory and the ambiguities associated with different 
UV completions are presented in Section \ref{sec:HIrem}. The effect of these completions  on the renormalization 
group enhanced potential is discussed in Section \ref{sec:effpot}. We consider two alternative 
scenarios and analyze their impact on the inflationary observables and on the complete power spectrum. The results are presented in 
Section \ref{sec:spectra} and  \ref{sec:spectra2}. Section \ref{sec:conclusions} contains the conclusions.

%%%%%%%%%%%%%%%%%%%%%%%%%%%%%
\section{Higgs inflation in the Einstein frame}\label{sec:HI}
%%%%%%%%%%%%%%%%%%%%%%%%%%%%%

The nonlinearities generated by the non-minimal coupling $\xi$ in Eq.~\eqref{lagr} can be transferred to the matter sector by performing a Weyl rescaling of the metric
$g_{\mu\nu}\rightarrow \Omega^2(h)g_{\mu\nu}$ with conformal factor $\Omega^2(h)=1+\xi h^2/M_P^2$. The resulting 
Einstein-frame action reads 
\begin{equation}\label{lagr2}
S= \int d^4x \sqrt{-g}\left[ \frac{M_P^2}{2}   R -\frac{1}{2}K(h)\left(\partial h\right)^2- V(h)\right]\,,
\end{equation}
with
\begin{equation}
  K(h)= \frac{\Omega^2(h)+6\xi^2h^2/M_P^2}{\Omega^4(h)}\,,
\end{equation}
the non-homogeneous part of the Ricci scalar transformation \cite{Maeda:1988ab} and 
\begin{equation}\label{potE}
V(h)=\frac{U(h)}{\Omega^4(h)} \,,
\end{equation}
the rescaled potential. In these variables, the dynamical information of the model 
is partially encoded in the non-canonical kinetic term. In order to compute the inflationary observables using the 
standard slow-roll techniques, it is convenient to perform an additional field redefinition 
\begin{equation}\label{dphidh}
\frac{d\phi}{dh}=\sqrt{K(h)}\,,
\end{equation}
to \textit{canonically normalize} the Higgs field. Although an exact analytical solution to this equation exists 
\cite{GarciaBellido:2008ab}, the result is not very enlightening. For the purposes of this paper, it will be enough to consider the approximate solution 
\begin{equation}\label{phih}
\phi=\begin{cases} 
   h &  \hspace{2mm} \textrm{for} \hspace{2mm} h<X_{\rm cr}\,,  \\
\sqrt{\frac{3}{2}} M_P \log \Omega^2(h) & \hspace{2mm} \textrm{for} \hspace{2mm} h>X_{\rm cr}\,,
   \end{cases}
\end{equation}
with
\begin{equation}
 X_{\rm cr}\equiv\sqrt{\frac{2}{3}} \frac{M_P}{\xi}\,,
\end{equation}
the critical value separating the low- and high-field regimes.  Using Eqs.~\eqref{potE} and \eqref{phih}, the Einstein-frame potential at $\phi\gg v_{\rm EW}$ becomes 
 \begin{equation}\label{Vtree}
 V(\phi)= \frac{\lambda}{4}F^4(\phi)\,,
 \end{equation}
with
 \begin{equation}\label{Ftree}
 F(\phi)\simeq 
 %\frac{h(\phi)}{\Omega(\phi)}=
\begin{cases} 
      \phi &  \hspace{2mm} \textrm{for} \hspace{2mm} \phi< X_{\rm cr}\,, \\
\frac{M_P}{\sqrt{\xi}} \left(1-e^{-\sqrt{\frac{2}{3}}\frac{\phi}{M_P}}\right)^{\frac12} 
 &  \hspace{2mm} \textrm{for} \hspace{2mm}\phi> X_{\rm cr}\,.
   \end{cases}
 \end{equation}
 Note that for $\phi\gg \sqrt{3/2}M_P$ the potential becomes exponentially flat. This asymptotic shift-symmetry is the 
 Einstein-frame manifestation of the approximate scale-invariance of Eq.~\eqref{lagr} at $h\gg M_P/\sqrt{\xi}$. The
 transition to the Einstein frame is indeed analogous to the spontaneous breaking of the symmetry. The field $\phi$ 
 can be interpreted as the pseudo-Goldstone boson of (non-linearly realized) scale-invariance \cite{Csaki:2014bua}.

For a \textit{canonically normalized} field $\phi$, the statistical information of the primordial perturbations generated 
during inflation is dominantly encoded in the two-point correlation function, or equivalently in its Fourier-transform, 
the power spectrum $\cal P_{\cal R}$. To the lowest  order in the slow-roll approximation, the amplitude of the
power spectrum reads 
\begin{equation}
 \mathcal{P}^*_{\cal R}\simeq \frac{1}{24\pi^2 \epsilon_*}\frac{V_*}{M_P^4}\,,
\end{equation}
with $\epsilon$ the first slow-roll parameter and the stars  denoting the evaluation of the associated quantities 
at horizon crossing, i.e. when $k=aH$. The normalization of the power spectrum at large scales \cite{Ade:2015lrj}
\begin{equation}\label{COBE}
\log (10^{10} \mathcal{P}^* _{\cal R}) \simeq 3.094 \pm 0.034\,,
\end{equation} fixes the amplitude of the inflationary potential and the tree-level relation between the Higgs 
self-coupling $\lambda$ and the non-minimal coupling $\xi$
\begin{equation}
\frac{\lambda}{\xi^2}\simeq 4\times 10^{-10}\,.
\end{equation}
 Note that any variation of $\lambda$ can be compensated for by a proper 
 change of $\xi$. Evaluating the slow-roll parameters $\epsilon$ and $\eta$ at horizon exit,
% \begin{equation}
% \epsilon_*=\left(\right)\simeq \frac{3}{4N^2}\,,\hspace{5mm} \eta_*=-\frac{1}{N}\,,
% \end{equation}
one obtains the following approximate expressions for the spectral tilt of scalar perturbations $n_s$ and the 
tensor-to-scalar ratio $r$ \cite{Bezrukov:2007ep}
 \begin{align}
 &n_s=1+2\eta_*-6\epsilon_* \simeq 1-\frac{2}{N}\,, \label{nsuni}\\ &r=16\epsilon_*\simeq \frac{12}{N^2}\,,\label{runi}
 \end{align}
 with $N$ the number of e-folds before the end of inflation. The precise value of $N$ in these equations 
 depends on the postinflationary history of the Universe and in particular on the duration of the reheating stage. Following
 the estimates of Refs.~\cite{GarciaBellido:2008ab,Bezrukov:2008ut,Repond:2016sol}, we will take $N\simeq 60$.  Numerically, 
 this gives rise to a spectral tilt in excellent agreement with the latest CMB results and a small tensor-to-scalar ratio
 \begin{equation}\label{nsrtree}
 n_s \simeq0.966\,,\hspace{10mm} r\simeq0.0033\,.
 \end{equation}

%%%%%%%%%%%%%%%%%%%%%%%%%%%%%
\section{Higgs inflation renormalization}\label{sec:HIrem} 
%%%%%%%%%%%%%%%%%%%%%%%%%%%%%
The non-minimal coupling between the SM Higgs and gravity translates into a 
non-renormalizable Einstein-frame potential \eqref{Vtree}. This lack of renormalizability also permeates
the interactions of the Higgs boson with the SM particles.

Higgs inflation should be interpreted as an EFT to be supplemented by a particular
set of higher-dimensional operators. In the absence of a UV completion of the SM coupled
to gravity, the choice of these higher-dimensional operators can only be based on the assumption of symmetries and/or on 
the self-consistency of the procedure. 

The minimal set of operators needed to make the theory finite at every 
order in perturbation theory follows from the theory itself~\cite{Bezrukov:2010jz,Bezrukov:2014ipa}. These operators 
are generated by radiative corrections and constitute a particular subset of a more general class of operators respecting 
the asymptotic scale-invariance of the tree-level action at $h\gg M_P/\sqrt{\xi}$, or equivalently, the shift-symmetry of the 
Einstein-frame potential \eqref{Vtree}.   The most important contributions are associated with the Higgs and 
top quark interactions. The relevant part of the Einstein-frame Lagrangian reads 
\begin{equation}
  \label{Ltree}
  \cL = -\frac{(\partial\phi)^2}{2} - \frac{\lambda}{4} F^4(\phi)
    + i \bar\psi_t\slashed\partial\psi_t - \frac{y_t}{\sqrt{2}} F(\phi) \bar\psi_t\psi_t\,,
\end{equation}
with $y_t$ the top Yukawa coupling. As usual, the loop divergences generated by this tree-level  Lagrangian can be eliminated by introducing 
a proper set of counterterms. A regularization scheme, that fits well with the asymptotic symmetries of Eq.~\eqref{lagr} 
at $h>M_P/\sqrt{\xi}$, is dimensional regularization with an asymptotically scale-invariant subtraction point \cite{Bezrukov:2009db}
\begin{equation}
\mu^2\propto M_P^2+\xi h^2\,.  
\end{equation}
Note that this is a Jordan-frame definition. For the results associated to other non-scale invariant prescriptions in this frame 
see Refs.~\cite{Barvinsky:2008ia, DeSimone:2008ei,Barvinsky:2009fy,Barvinsky:2009ii}.

On general grounds, a counterterm in dimensional regularization
contains a pole in $\epsilon=(4-d)/2$ (with $d$ the fractional spacetime dimension) and a finite part $\delta{\cal L}$
\begin{equation}\label{dimreg}
\Delta{\cal L}=\frac{A}{\epsilon} +\delta{\cal L}\,.
\end{equation}
The coefficients of the poles are chosen to cancel the divergences generated by loop diagrams. Once
the divergent parts are subtracted we are left only with the finite terms. At 1-loop, they read~\cite{Bezrukov:2014ipa}
\begin{align}\label{counterL}
  \delta\cL^{F}_\text{1}& = \left[\delta\lambda_a \left(F'^2+\frac{1}{3}F''F\right)^2 -\delta\lambda_b\right] F^4\,, \\
  \delta\cL^{\psi}_\text{1} &=\left[\delta y_{a} F'^2F  +\delta y_{b} F''(F^4)''\right]\bar\psi\psi\,,\label{counteryt}
\end{align}
with the primes denoting derivatives with respect to the scalar field $\phi$.
The coefficients $\delta\lambda_a,\delta\lambda_b,\delta y_{a}$ and $\delta y_{b}$ in these equations should be understood as
remnants of a given UV completion and cannot be determined from the EFT itself 
\cite{Bezrukov:2010jz,Bezrukov:2014ipa,Burgess:2014lza}. 

Note that the functional dependence of Eqs.~\eqref{counterL} and \eqref{counteryt} 
differs from that in the tree-level Lagrangian. While the finite part $\delta\lambda_b$ can be reabsorbed 
into the definition of the Higgs-self coupling, the consistency of the renormalization procedure requires
promoting $\delta\lambda_a, \delta y_{a}$ 
and $\delta y_{b}$ to new coupling constants with their own renormalization group equations. The inclusion of the 
associated counterterms in the tree-level Lagrangian and the re-evaluation of radiative corrections generates  
new contributions on top of the original one-loop result. 
As before, these finite parts should be promoted to 
new coupling constants with additional renormalization group equations. This iterative scheme gives rise 
to a renormalized Lagrangian containing a tower of higher-dimensional operators. The presence of these 
operators has far-reaching consequences. In particular, the masses of the SM  particles at the electroweak scale \textit{cannot} 
be unequivocally related to their inflationary values \textit{without the precise knowledge of the UV completion} 
\cite{Bezrukov:2014ipa} (see also Refs.~\cite{Burgess:2014lza,Hertzberg:2011rc}).
  
In Ref.~\cite{Bezrukov:2014ipa}, it was assumed that the finite parts appearing at every order in perturbation 
theory were small and of the same order as the loops generating them. In practice, this 
assumption allows to safely truncate the aforementioned set of higher-dimensional operators. Note, however, that 
the functional form of the effective action at small and large field values is almost insensitive to this 
assumption~\cite{Bezrukov:2010jz,Bezrukov:2014ipa}. In particular, all threshold effects (independent of their order) maintain the quartic behavior of the Higgs 
potential at $\phi\ll X_{\rm cr}$ and the asymptotic flatness at $\phi\gg\sqrt{3/2}\,M_P$ \cite{Bezrukov:2014ipa}. The finite 
parts can only affect the \textit{transition region}\footnote{Equivalently, for the original Jordan-frame field $h$, the transition
  region is $X_{\rm cr}< h < M_P/\sqrt{\xi}$.} $X_{\rm cr}<\phi<\sqrt{3/2} \,M_P$. This can easily be understood by considering the 
structure of Eq.~\eqref{Ftree}. 
For small field values the function $F(\phi)$ in Eq.~\eqref{Vtree} becomes approximately linear, $F \approx \phi$ and the conformal factor equals one,
$\Omega \approx 1$, up to  highly-suppressed  corrections of the order $\mathcal{O}(\phi/(M_P/\xi))$. 
The quartic behavior of the potential at $\phi< X_{\rm cr}$ is therefore approximately maintained.
Equivalently, the Jordan-frame cutoff scale \eqref{Lcut} becomes approximately $\Lambda(h)\approx M_P/\xi$ at small $h$. All operators of the shape \eqref{genope} with $n>4$ are supressed by a scale that significantly exceeds 
the Higgs vacuum expectation value, $M_P/\xi\gg v_\text{EW}$, and the energy scale of any present collider physics 
experiment \cite{Barbon:2009ya,Burgess:2009ea,Burgess:2010zq,Bezrukov:2010jz}.
While the above argument only considers the Higgs sector, the same suppression applies to higher-order operators involving the SM fermions. 
The finite parts of the coefficients of \textit{all} dimension four operators can be reabsorbed into the definition of the low energy couplings, as is usually done in 
renormalizable field theories.  

However some of these contributions dynamically vanish in the intermediate 
region $X_{\rm cr}<\phi<\sqrt{3/2} \,M_P$, where the function $F(\phi)$ evolves towards the constant value $F_\infty = M_P/\sqrt{\xi}$ 
and the \textit{full} set of higher-dimensional operators becomes exponentially suppressed. The same argument can be extended 
to other operators beyond the minimal set. Indeed, the Einstein-frame version of the asymptotically scale-invariant 
operators \eqref{genope} 
\begin{equation}\label{genopeE}
\frac{1}{\Omega^4}\frac{c_n{\cal O}_{n}(h)}{\left[\Lambda(h)\right]^{n-4}}\,.
\end{equation}
become asymptotically shift-symmetric in the large field regime.

%%%%%%%%%%%%%%%%%%

%%%%%%%%%%%%%%%%%%%%%%%%%%%%%
\section{The effective potential}\label{sec:effpot} 
%%%%%%%%%%%%%%%%%%%%%%%%%%%%%

The threshold effects in the \textit{transition region} $X_{\rm cr}<\phi<\sqrt{3/2} \,M_P$ modify the running  of 
the SM couplings (see Refs.~\cite{Bezrukov:2012sa,Degrassi:2012ry,Buttazzo:2013uya} for the pure SM running and 
Refs.~\cite{George:2015nza,Bezrukov:2009db,Bezrukov:2008ej} for the running in Higgs inflation). This modification can be incorporated in the action by replacing the tree-level couplings by their running values in the presence of threshold corrections \cite{Bezrukov:2014bra,Bezrukov:2014ipa}. The renormalization group enhanced (RGE) potential at 
$\phi\gg v_{\rm EW}$ takes the approximate form  
\begin{equation}\label{Veff}
V(\phi)=\frac{\lambda (\phi)}{4}F^4(\phi)\,,
\end{equation}
with
\begin{equation}
\label{Lrun}
\lambda(\phi) =\lambda_0+b \log^2 \left(\frac{m_t(\phi)}{q}\right)+\delta\Lambda(\phi)\,.
\end{equation}
The quantities $\lambda_0$ and $b$ in this expression should be understood as functions 
of the \textit{Einstein-frame} top quark mass $m_t(\phi) = \frac{y_t}{\sqrt{2}} F(\phi)$ and the
Higgs mass \textit{at the inflationary scale} \cite{Bezrukov:2014bra}.  The function $\delta\Lambda(\phi)$ relates $\lambda_0$ to its low energy counterpart, i.e. it parametrizes the effect 
of threshold corrections.
 For $\delta\Lambda(\phi) = 0$ the parameter $q$ marks the energy scale at which the beta function
 for the Higgs self-coupling $\lambda(\phi)$ vanishes. 
 The parameter $b$ corresponds to the derivative of the beta function at that scale, with 
 $b\simeq2.3\times10^{-5}$ for the SM renormalization group.
The choice\footnote{ This corresponds to the prescription I of Ref.~\cite{Bezrukov:2009db}.}
\begin{equation}\label{muscale}
\frac{m_t}{q} \equiv \frac{\sqrt{\xi}F(\phi)}{\kappa M_P}\,,
\end{equation}
respects the asymptotic symmetry of the scaling frame \eqref{lagr} at $h\gg M_P/\sqrt{\xi}$ and the corresponding shift-symmetry of the Einstein-frame action at $\phi>\sqrt{3/2}M_P$. Here we have defined an effective parameter $\kappa$, which depends on the scale $q$, the non-minimal coupling $\xi$ and the details of the one-loop potential calculation.  This parameter is convenient for the computation of the inflationary observables and can be easily translated into more ``physical'' quantities via Eq.~\eqref{muscale}.  Note, however, that due to the presence of threshold corrections, these  ``physical'' quantities are not directly the parameters obtained from the renormalization group running of the SM at low energies.
  
In the lack of a UV completion able to determine the coefficients of higher-dimensional operators, the precise shape 
of the potential in the \textit{transition region}  $X_{\rm cr}<\phi<\sqrt{3/2} \,M_P$ cannot be unequivocally determined. 
In Ref.~\cite{Bezrukov:2014bra} (see also Refs.~\cite{Fumagalli:2016lls,Enckell:2016xse}), the threshold 
correction $\delta \Lambda(\phi)$ was taken to be completely \textit{field-independent} (or \textit{instantaneous}).
In this limit, Eq.~\eqref{Veff} becomes \cite{Bezrukov:2014bra} 
\begin{equation}\label{Veff2}
V(\phi)=\frac{1}{4}\left(\lambda_0+b \log^2 \left(\frac{\sqrt{\xi}F(\phi)}{\kappa M_P}\right)\right)F^4(\phi)\,.
\end{equation}
This RGE potential displays some distinct features with respect to the tree-level potential. In particular, 
it allows for the appearance of an exact inflection point along the inflationary trajectory for particular values of the 
model parameters, namely for
\begin{equation}
 \lambda_0 \, = \frac{b}{16\kappa}\,.
\end{equation}
The position of the inflection point 
\begin{equation}\label{phiinf}
\phi_{\rm inf}=\sqrt{\frac{3}{2}}\log\left(\frac{\sqrt{e}}{\sqrt{e}-1}\right)M_P\,,
\end{equation}
turns out to be numerically close to the value $\sqrt{3/2}{M_P}$,
where the function $F(\phi)$ in Eq.~\eqref{Ftree} becomes effectively field independent.

  %%%%%%%%%%%%%%%%%%%%%%%%%%%%%  %%%%%%%%%%%%%%%%%%%%%%%%%%%%%
\begin{figure}
\centering
  \includegraphics[scale=0.283]{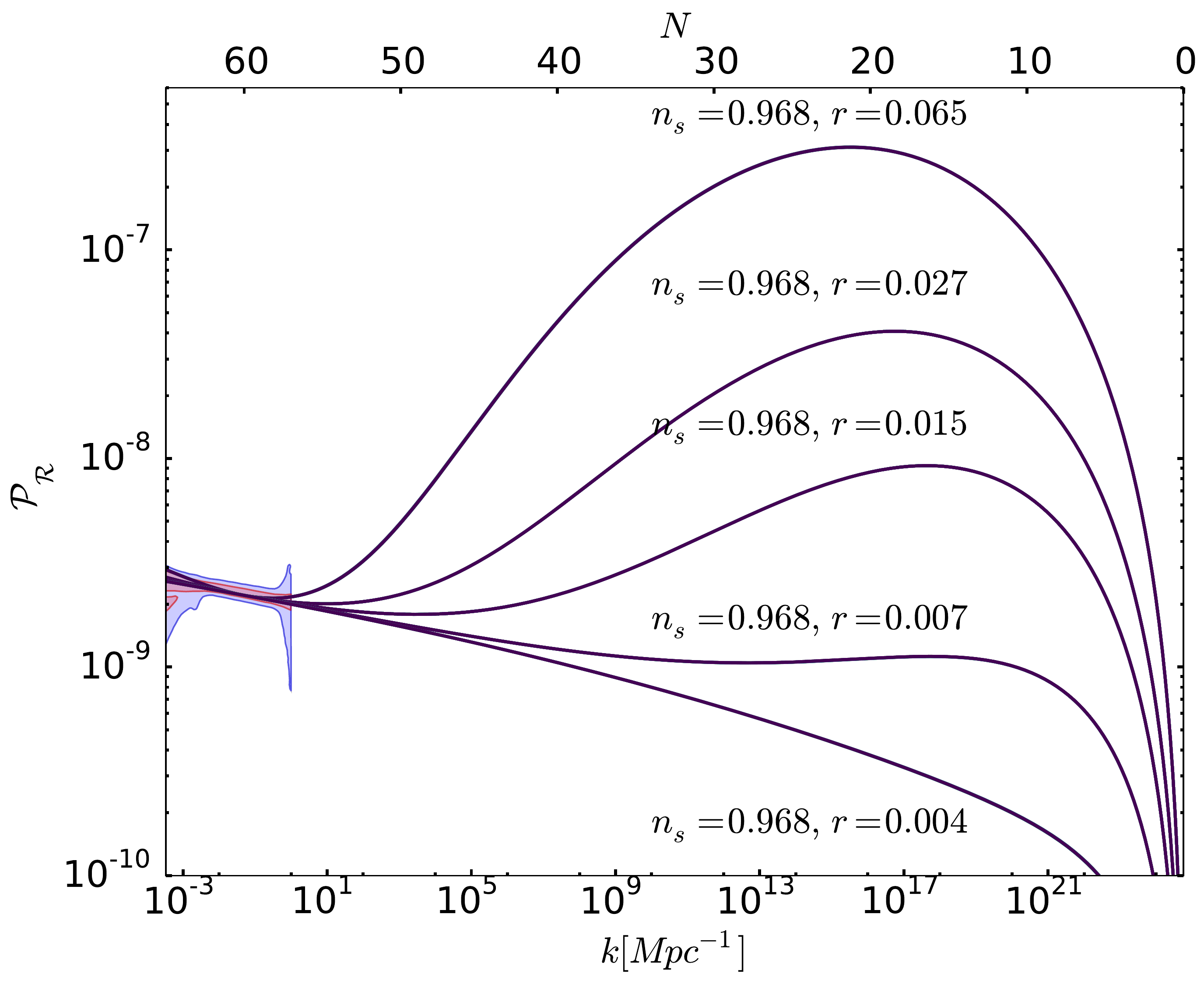}
\caption{The inflationary power spectrum ${\cal P}_{\cal R}$ for the \textit{instantaneous threshold} scenario \eqref{Veff2} as a function of momenta 
and the number of e-folds before the end of inflation. For multiple values of $\kappa$, the non-minimal coupling 
$\xi$ was varied to fix $n_s=0.968$. The lower curve corresponds to the \textit{universal/non-critical regime} with 
$\lambda_0\gg b/16\kappa$. The upper curves stand for different realizations of the \textit{critical regime}, all 
with values of $r$ within the Planck 95\% C.L. contour. The shaded regions at low $k$ mark the 68\% and 95\% C.L. constraints 
on the power spectrum \cite{Ade:2015lrj}.}\label{fig:spectra}
\end{figure}
  %%%%%%%%%%%%%%%%%%%%%%%%%%%%%  %%%%%%%%%%%%the%%%%%%%%%%%%%%%%%
    %%%%%%%%%%%%%%%%%%%%%%%%%%%%%  %%%%%%%%%%%%%%%%%%%%%%%%%%%%%
\begin{figure}
\vspace{8pt}
\centering
  \includegraphics[scale=0.29]{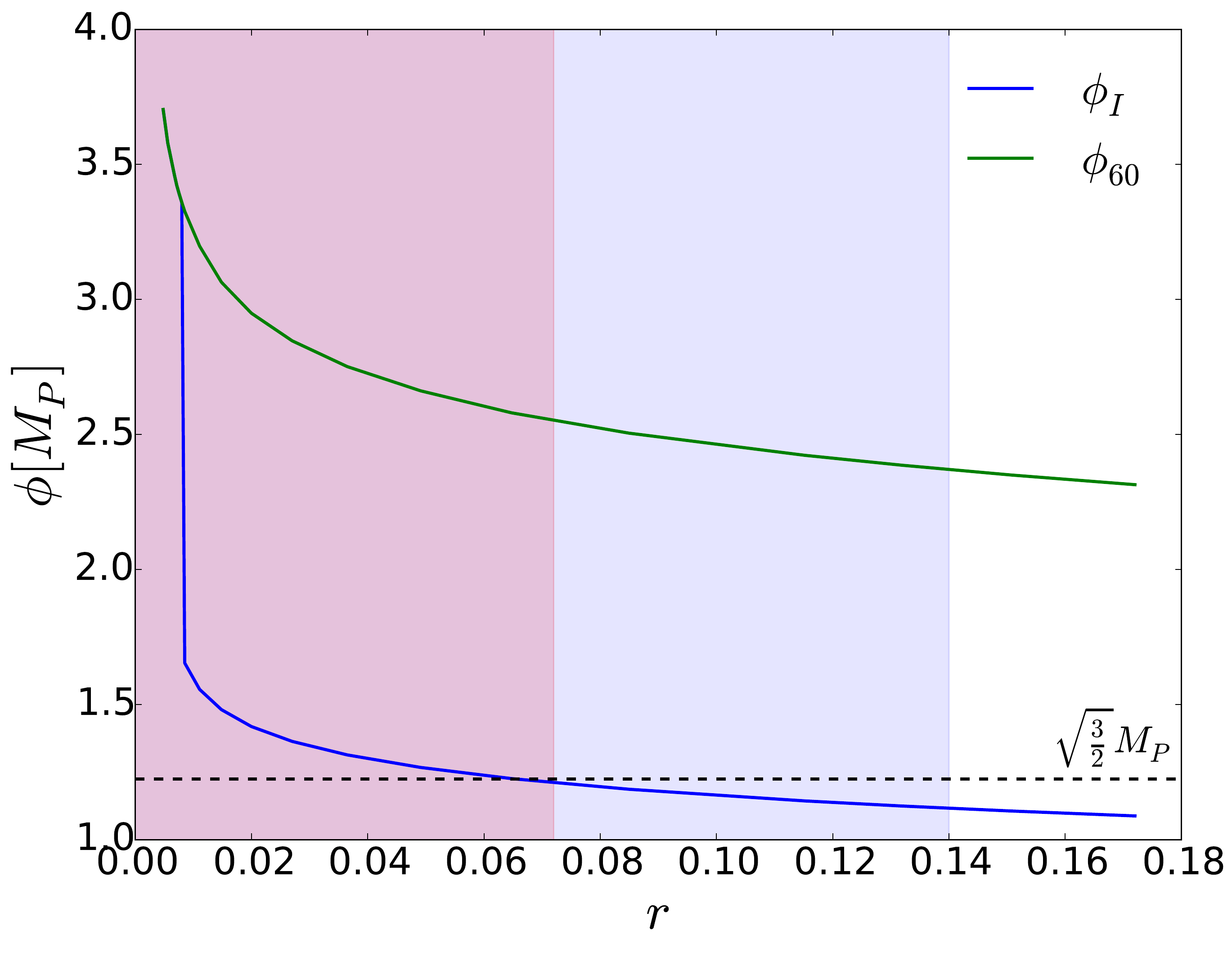}
\caption{The position of the \textit{quasi}-inflection point for a fiducial spectral tilt $n_s=0.968$ and 
varying tensor-to-scalar ratio $r$. The shaded regions mark the 68\% and 95\% Planck constraints on $r$ \cite{Ade:2015lrj}. The sharp transition at $r\simeq0.008$ signals the appearance of the \textit{quasi}-inflection point in this particular 
example.}\label{fig:infpoint}
\end{figure}
  %%%%%%%%%%%%%%%%%%%%%%%%%%%%%  %%%%%%%%%%%%%%%%%%%%%%%%%%%%%  
   
The power spectrum following from Eq.~\eqref{Veff2} is shown in Fig.~\ref{fig:spectra} for different 
  values of $\kappa$ and $\xi$. Due to the complicated shape of this RGE potential, we rely on numerical methods. 
  Using the standard slow-roll approximation, for each pair of $\kappa$ and $\xi$, we obtain the inflaton value at the onset of the inflationary regime, assuming $60$ e-folds of inflation. 
  We then adjust the  constant offset $\lambda_0$ to get the right normalization \eqref{COBE} and compute the inflationary observables and the power spectrum along the inflationary trajectory. 
  We also require the inflationary potential to be monotonic during the whole inflationary regime. 
  In monotonic potentials that are extremely flat one can enter an era of \textit{ultra slow-roll} 
  \cite{Kannike:2017bxn,Germani:2017bcs}, in 
  which the slow-roll approximation breaks down. In this regime stochastic effects become important 
  \cite{Starobinsky:1994bd,Vennin:2015hra,Pattison:2017mbe}. We explicitly confirmed that we do not need to consider these effects 
  by numerically integrating the equations of motion and checking that the Hubble flow parameters
  $\bar{\epsilon} = - \frac{\dot{H}}{H^2}$ and $\bar{\epsilon}_2 \equiv \frac{\dot{\epsilon}}{\epsilon H}$ remain much smaller 
  than one along the whole inflationary trajectory. In all the scenarios considered in this paper
  the standard slow-roll approximation always remains valid.
  By varying the parameter $b$ in the interval $0.9 - 2.3 \times 10^{-5}$, 
  we confirmed that the inflationary predictions  are independent of its precise value.  
  Thus, the difference between the exact renormalization group evolution near the scale $q$ and the pure SM running one is numerically not important. 
  For illustration purposes,  we choose  the SM value $b=2.3 \times 10^{-5}$. As shown in Fig.~\ref{fig:spectra}, we can clearly distinguish two regimes:
\begin{enumerate}[i)]
\item {\it Universal/Non-critical regime}: For $\lambda_0 \gg b/16\kappa$, the effective potential 
becomes almost insensitive to the logarithmic term in Eq.~\eqref{Veff2}. 
The value of the first slow-roll parameter $\epsilon$ increases monotonically towards the end of inflation. This is 
reflected in the featureless shape of the power spectrum in this regime. The spectral tilt and the tensor-to-scalar 
ratio coincide with their tree-level values \eqref{nsrtree}. Our results also agree with those in 
Refs.~\cite{Fumagalli:2016lls,Enckell:2016xse,Bezrukov:2014bra}.

\item{\it Critical regime}: If $\lambda_0 \simeq  b/16\kappa$, the effective potential develops a 
\textit{quasi}-inflection point around which the inflaton evolves very slowly 
as compared to the rolling in other parts of the inflationary trajectory. This behavior 
modifies the inflationary observables \cite{Bezrukov:2014bra,Hamada:2014iga,BenDayan:2009kv,Hotchkiss:2011gz}. In particular, the tensor-to-scalar ratio $r$ can become comparatively large, $r\simeq {\cal O}(10^{-2}-10^{-1})$. On top of the 
large-scale modifications of the spectrum, the non-monotonic evolution of $\epsilon$ generates a bump at small and 
intermediate scales. At fixed spectral tilt $n_s$, the amplitude and width of the peak are correlated with the 
tensor-to-scalar ratio  $r$. Larger values of $r$ give rise 
to higher and wider peaks. The maximum values of the spectrum compatible  with the 68\% and
95\% C.L Planck results are ${\cal P}^{\rm max}_{\cal R}\simeq 3.7\times 10^{-7}$ and 
 ${\cal P}^{\rm max}_{\cal R}\simeq 1.6\times 10^{-6}$, respectively. Note, however, that in the presence of the bump the
 global behavior of the spectrum cannot be accurately described by a simple
 expansion in terms of the spectral-tilt and its running. Although the direct comparison of  CMB data 
 with the primordial spectrum at large scales displays a reasonable consistency  (cf. Fig. \ref{fig:spectra}), it
 would be interesting to eventually perform a complete fit of the spectrum, as suggested in Ref.~\cite{Lesgourgues:2007gp}. 
\end{enumerate}

The field value $\phi_I$ associated with the minimum of the slow-roll parameter $\epsilon$ for  a
fiducial spectral tilt $n_s=0.968$ and varying tensor-to-scalar ratio $r$ is shown in Fig.~\ref{fig:infpoint}. The shaded regions stand
for  the 68\% and 95\% Planck constraints on $r$. For 
the universal regime with monotonic slow-roll parameter $\epsilon$, the value of $\phi_I$ coincides with the value of the 
field at the onset of the inflationary regime, $\phi_{60}$. The sharp transition at $r\simeq 0.008$ signals the 
beginning of the critical regime in this particular example. Beyond this point, the quantity $\phi_I$ approximately coincides with the position of the \textit{quasi}-inflection point. The larger the tensor-to-scalar ratio, 
the closer the \textit{quasi}-inflection point is to the region $X_{\rm cr}<\phi<\sqrt{3/2}M_P$, where threshold effects are 
expected to be relevant.

\section{Threshold corrections and inflationary observables}\label{sec:spectra} 
 
 In this section, we analyze the impact of \textit{non-instantaneous} threshold corrections in the inflationary observables.   We consider two different scenarios: 1) threshold effects consistent with the truncation of the renormalization group  equations at 1-loop and  2) collective threshold effects parametrized by a smoothly interpolating function between   the low- and high-energy regimes. 

\subsection{1-loop threshold scenario}\label{sec:1loop}

This scenario follows the lines of Ref.~\cite{Bezrukov:2014ipa}. In particular, we will assume that 
the finite parts appearing at every order in perturbation theory are small and of the same order as 
 the loops producing them. Using this hierarchy to consistently truncate the series of higher-dimensional 
 operators at first order in perturbation theory, and taking into account Eq.~\eqref{counterL}, we will 
 parametrize the effect of the 1-loop jumps by an effective function\footnote{As explained in Section \ref{sec:HIrem}  the coefficient $\delta\lambda_a$ in Eq.~\eqref{thresh1loop} should be promoted to 
 a new coupling constant with its own renormalization group equation. Here we explicitly neglect the
 running of $\delta\lambda_a$ in the transition 
 region  $X_{\rm cr}<\phi<\sqrt{3/2}M_P$ and postpone the computation of the associated beta function to a future work.}
 \begin{equation}\label{thresh1loop}
\delta\Lambda(\phi)=\delta\lambda_a \left(F'^2+\frac{1}{3}F''F\right)^2\,.
\end{equation}
     %%%%%%%%%%%%%%%%%%%%%%%%%%%%%  %%%%%%%%%%%%%%%%%%%%%%%%%%%%%
\begin{figure}
 \begin{flushleft}
  \includegraphics[scale=0.83]{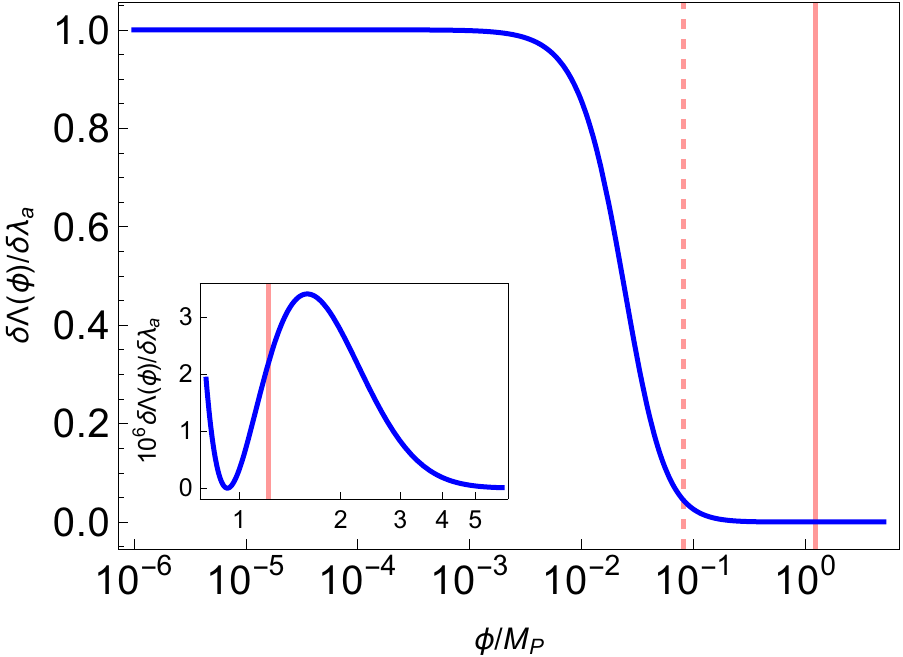}
\caption{The 1-loop threshold correction \eqref{thresh1loop} for $\xi=10$. This particular value corresponds to the typical magnitude of the non-minimal coupling $\xi$ in the \textit{critical regime} and is chosen for illustration purposes only. The vertical lines correspond 
to $\phi=X_{\rm cr}$ (dashed red) and $\phi=\sqrt{3/2}M_P$ (solid red). In the inset we show a rescaled version of 
$\delta\Lambda(\phi)$ resolving the small bump generated at $\phi>\sqrt{3/2}M_P$ by the sign flip in $F''$.}\label{fig:1loopthres}
 \end{flushleft}
\end{figure}
  %%%%%%%%%%%%%%%%%%%%%%%%%%%%%  %%%%%%%%%%%%%%%%%%%%%%%%%%%%%

    %%%%%%%%%%%%%%%%%%%%%%%%%%%%%  %%%%%%%%%%%%%%%%%%%%%%%%%%%%%
\begin{figure}
\centering
  \includegraphics[scale=0.245]{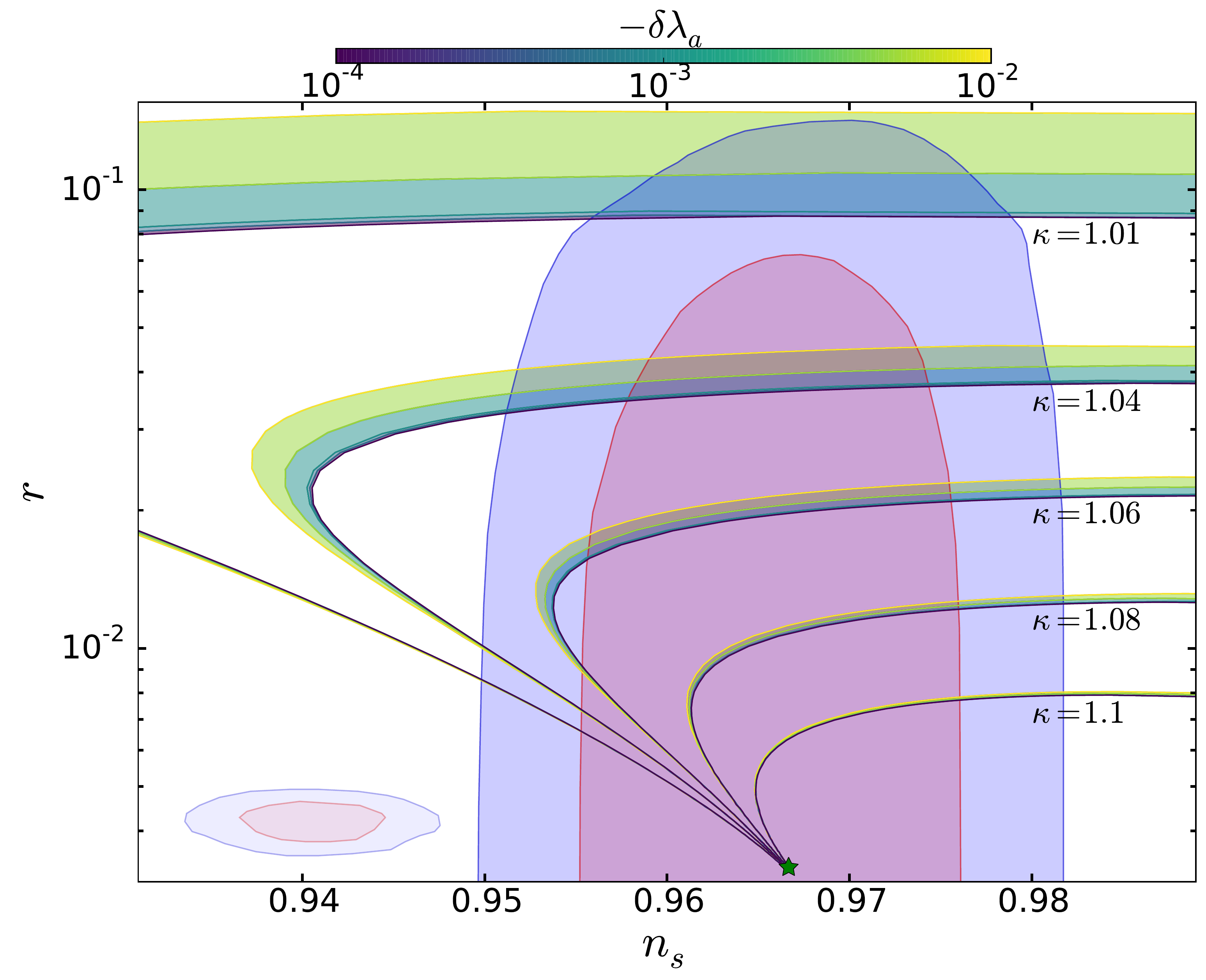}
\caption{The ($n_s,r$) plane for the \textit{1-loop threshold} scenario. Different contours correspond
to different values of $\delta\lambda_a$ with fixed $\kappa$.  Along the lines of constant $\kappa$, we vary the non-minimal coupling $\xi$ in the range $10-100$. The shaded 
regions mark the latest Planck  
constraints at 68\% and 95\% C.L. \cite{Ade:2015lrj} while the small contour in the bottom left corner shows the projected sensitivity of the 
CORE mission  \cite{Finelli:2016cyd}. The star indicates the universal values given by Eqs.~\eqref{nsuni} and
\eqref{runi}, corresponding to ${\cal O}(100)$ values of $\xi$.  }\label{fig:nsvsr1loop}
\end{figure}
  %%%%%%%%%%%%%%%%%%%%%%%%%%%%%  %%%%%%%%%%%%%%%%%%%%%%%%%%%%%
The shape of this threshold correction is shown in Fig.~\ref{fig:1loopthres}. After a fast 
decay,  $\delta\Lambda(\phi)$ develops a small bump\footnote{This bump is due to a change of sign in $F''$  at $\phi>\sqrt{3/2}M_P$.} and approaches zero exponentially at large field values. 
The inflationary observables associated to this scenario are shown in Fig. \ref{fig:nsvsr1loop}  for different values 
of $\kappa$ and amplitudes $\delta\lambda_a$ in the range $-10^{-2}\leq \delta\lambda_a \leq -10^{-4}$. 
The choice $\delta\lambda_a\geq -10^{-2}$ is based on particle 
physics phenomenology and on our interest in the critical Higgs inflation scenario. \textit{At the inflationary scale} 
the Higgs self-coupling in the critical regime is of the order  of $\mathcal{O}(10^{-6})$. 
\textit{At the electroweak scale} the value of the Higgs self-coupling can be determined from experiment and its running 
computed using the SM renormalization group equations.  
A jump $\delta\lambda_a\sim {\cal O}(-10^{-2})$ connects the inflationary value of the Higgs self-coupling associated 
to the critical regime with 
values of the Higgs mass at the electroweak scale in good agreement with the latest LHC results
\cite{Bezrukov:2014ipa}.\footnote{See Ref.~\cite{Bezrukov:2014ina} for 
the uncertainties associated to the determination of the top Yukawa coupling from the Monte Carlo top quark mass.}
\textit{Within the critical regime}, larger jumps would be in conflict with 
 electroweak constraints (that is, correspond to too light Higgs masses). 
 On the contrary, sizable values of $\delta\lambda_a$, including $\delta\lambda_a\sim {\cal O}(-1)$, are
 allowed in the \textit{non-critical regime}, leading  to values of the Higgs self-coupling at inflation larger than $\mathcal{O}(10^{-6})$ \cite{Bezrukov:2014ipa}. 
 Note, however, that for $|\delta\lambda_a|\gtrsim\mathcal{O}(1)$ the perturbative computation, 
  and the associated truncation of the renormalization group equations leading to Eq.~\eqref{thresh1loop}, is 
  expected to break down.

Varying $\delta\lambda_a$, again we can distinguish two regimes:

\begin{enumerate}[i)]
\item {\it Universal/Non-critical regime}:
In the absence of a \textit{quasi}-inflection point able to account for a large number of e-folds, the value 
of the inflaton at horizon exit significantly exceeds the values affected by threshold corrections even if these are not completely instantaneous. 
In this regime, the inflationary observables are independent of the precise choice of $\delta\lambda_a$ and $\kappa$. 
Different values of $\delta\lambda_a$ can always be accomodated by different values of the non-minimal 
 coupling $\xi$ and the observables  approach the attractor values in Eq.~\eqref{nsrtree}. This is consistent with the results of Ref.~\cite{Fumagalli:2016lls, Enckell:2016xse}.
\item{\it Critical regime}:   Since the value of the tensor-to-scalar ratio at horizon exit  is related to the fraction of
e-folds accounted for in the \textit{quasi}-inflection point, small modifications of the potential around this point can
translate into sizable effects on the inflationary observables. 
For very small values of $\delta\lambda_a$, the slow-roll evolution of the inflaton field in the vicinity of the
\textit{quasi}-inflection point is not significantly affected. In this case, the predictions of Higgs inflation remain reasonably stable \textit{even within the critical regime}. This is due to the $1/\xi^2$ suppression of Eq.~\eqref{thresh1loop} and its rapid decay at $\phi\sim X_{\rm cr}$. Note also that 
the localized bump in Fig.~\ref{fig:1loopthres} does not give rise to any observable effect. Its relative size is so small,
that the bump is numerically irrelevant as compared to the exponential rising of the potential in the same field region.
The situation changes completely for larger values of $\delta\lambda_a$, i.e. when the contribution of $\delta \Lambda(\phi)$ 
at   the inflection point (\ref{phiinf}) becomes comparable with $\lambda_0$. As shown in Fig.~\ref{fig:nsvsr1loop}, the 
inflationary observables generically depend on the details of the UV completion. 

\end{enumerate}

\subsection{Collective threshold scenario}\label{sec:coll}

To illustrate the degeneracies between amplitude and rapidity of  threshold corrections, we will consider a purely phenomenological scenario. Rather than assuming a particular set of operators sharing the asymptotic shift-symmetry of the Einstein-frame tree-level action, we will parametrize  their \textit{collective effect} by a step-like function smoothly interpolating between the low and high-energy regimes, namely
\begin{equation}\label{thresh}
\delta\Lambda(\phi)=\delta\lambda\frac{\left(1-F^2/F^2_\infty\right)^4}{\left(1+\Delta\cdot 6\,\xi \,F^2/F^2_\infty\right)^2}\,,
\end{equation}
with $F_\infty=M_P/\sqrt{\xi}$ the asymptotic value of $F$ at $\phi\rightarrow\infty$. The parameters $\delta\lambda$ and $\Delta$ in this expression account for the amplitude and rapidity of the transition.   For $\Delta=1$, the shape of the correction coincides \textit{exactly} with the structure $(F')^4$, typically appearing in 1-loop expressions.
     %%%%%%%%%%%%%%%%%%%%%%%%%%%%%  %%%%%%%%%%%%%%%%%%%%%%%%%%%%%
\begin{figure}
 \begin{flushleft}
  \includegraphics[scale=0.83]{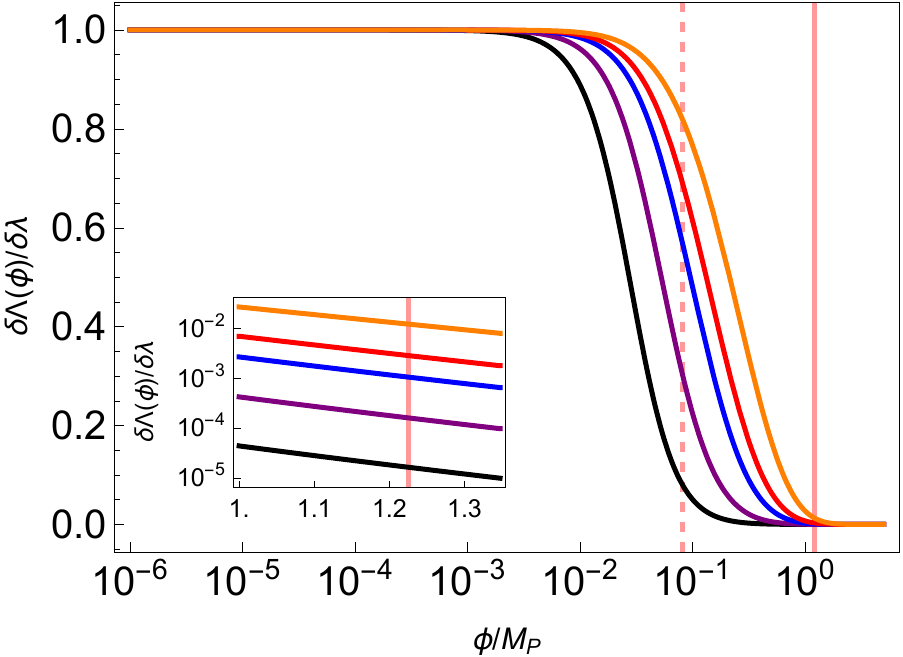}
\caption{The typical behavior of  the collective-threshold parametrization \eqref{thresh} for a fiducial value $\xi=10$ and $\Delta=1, 0.3, 0.1, 0.05$ and $0.01$ (starting from the lower curve). The vertical lines correspond 
to $\phi=X_{\rm cr}$ (dashed red) and $\phi=\sqrt{3/2}M_P$ (solid red). In the inset we show a rescaled version of 
$\delta\Lambda(\phi)$ resolving the vicinity of $\phi=\sqrt{3/2}M_P$.}\label{fig:thres}
 \end{flushleft}
\end{figure}
  %%%%%%%%%%%%%%%%%%%%%%%%%%%%%  %%%%%%%%%%%%%%%%%%%%%%%%%%%%%
    %%%%%%%%%%%%%%%%%%%%%%%%%%%%%  %%%%%%%%%%%%%%%%%%%%%%%%%%%%%
\begin{figure}
\centering
  \includegraphics[scale=0.248]{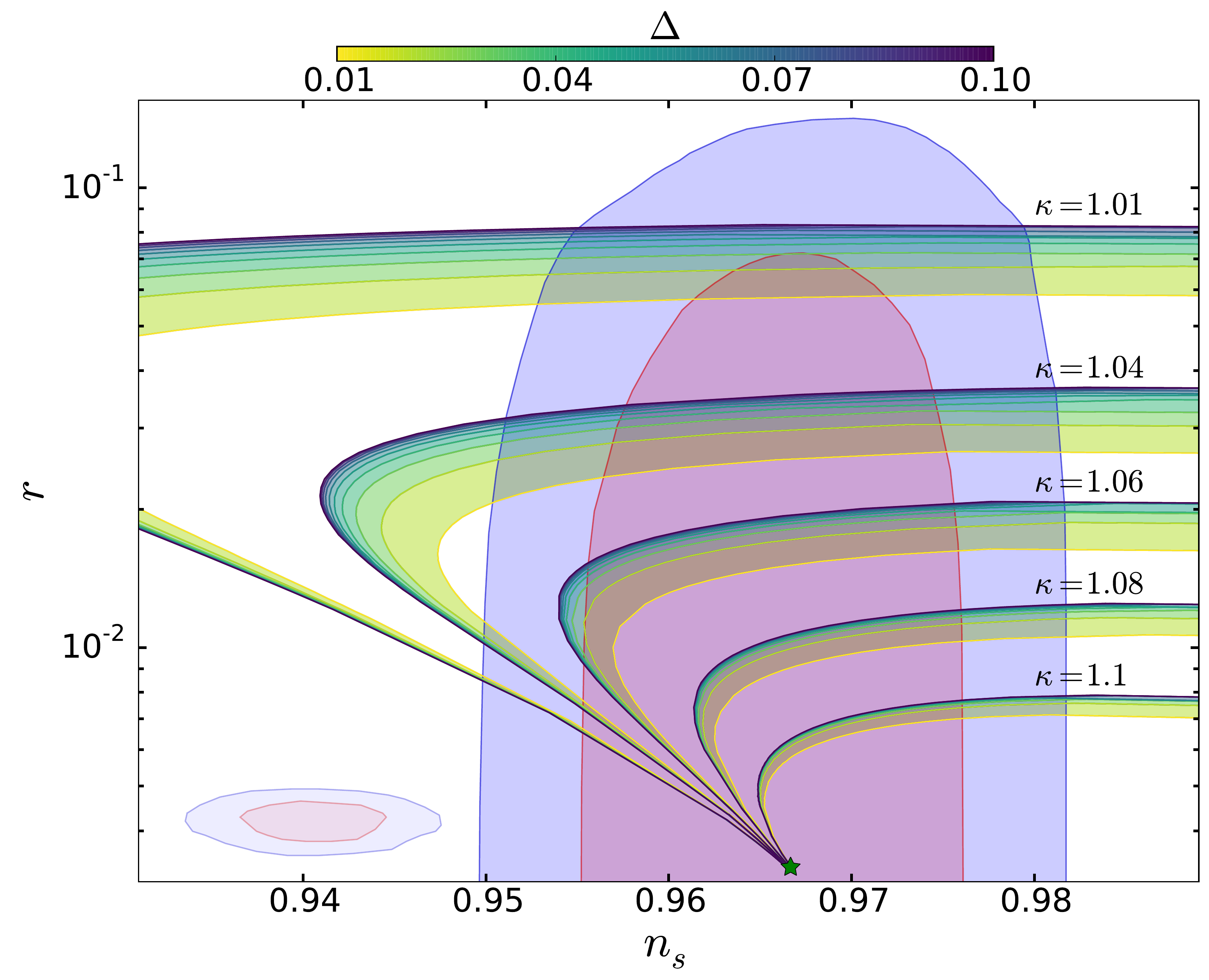}
\caption{The ($n_s,r$) plane for the \textit{collective threshold} scenario with $\delta\lambda= -1 \times 10^{-5}$ and its comparison 
with the latest Planck results at 68\% and 95\% C.L. \cite{Ade:2015lrj}.  Again, we vary $\xi$ in the range $10-100$ along lines of constant $\kappa$ . Different contours correspond
to different values of $\Delta$ with fixed $\kappa$. The contours in the bottom left corner show the projected sensitivity 
of the CORE mission at 68\% and 95\% C.L. \cite{Finelli:2016cyd}.}\label{fig:nsvsrjumps}
\end{figure}
  %%%%%%%%%%%%%%%%%%%%%%%%%%%%%  %%%%%%%%%%%%%%%%%%%%%%%%%%%%%
The variation of $\Delta$ is intended to encode effects that were explicitly ignored in the \textit{1-loop} scenario, such as the running of the finite parts or the constructive/destructive interference of additional operators if the 1-loop truncation is not assumed (see Fig.~\ref{fig:thres}). Note that Eq.~\eqref{thresh} is chosen for illustration purposes only. Its particular shape is not essential for the results presented here\footnote{Alternative interpolating functions should lead to qualitatively similar results.
}. Indeed, the precise way in which the ultraviolet regime is connected to $X_{\rm cr}$ is completely irrelevant 
for the determination of the inflationary observables. Only the variation of  $\delta \Lambda (\phi)$ in the vicinity
of the \textit{quasi}-inflection point is important. 

By varying $\Delta$ we explore the effect of changes in the rapidity of the transition on observables. 
For less rapid transitions even tiny jumps that did not play a role in the one-loop case can become relevant.
We highlight this feature by choosing a fiducial value of $\delta\lambda = -1 \times 10^{-5}$.
The modifications of the primordial power spectrum, the spectral tilt and the tensor-to-scalar ratio for
this fiducial value and finite values of $\Delta$ are presented in Fig. \ref{fig:nsvsrjumps}. This figure clearly 
illustrates the degeneracy between the amplitude of the jumps and the rapidity of the transition. 
Small jumps $\delta\lambda \simeq -1 \times 10^{-5}$ that \textit{were not} significantly modifying 
the inflationary observables in the \textit{1-loop scenario}  become relevant for slower transitions. 
If we had considered jumps of order $\delta\lambda \sim \mathcal{O}(-10^{-2})$, even small changes of 
order $\Delta \sim \mathcal{O}(10^{-1})$ in the transition rapidity would have significantly altered the observables.

\section{Threshold corrections and the primordial power spectrum}\label{sec:spectra2}

\begin{figure}
\centering
  \includegraphics[scale=0.248]{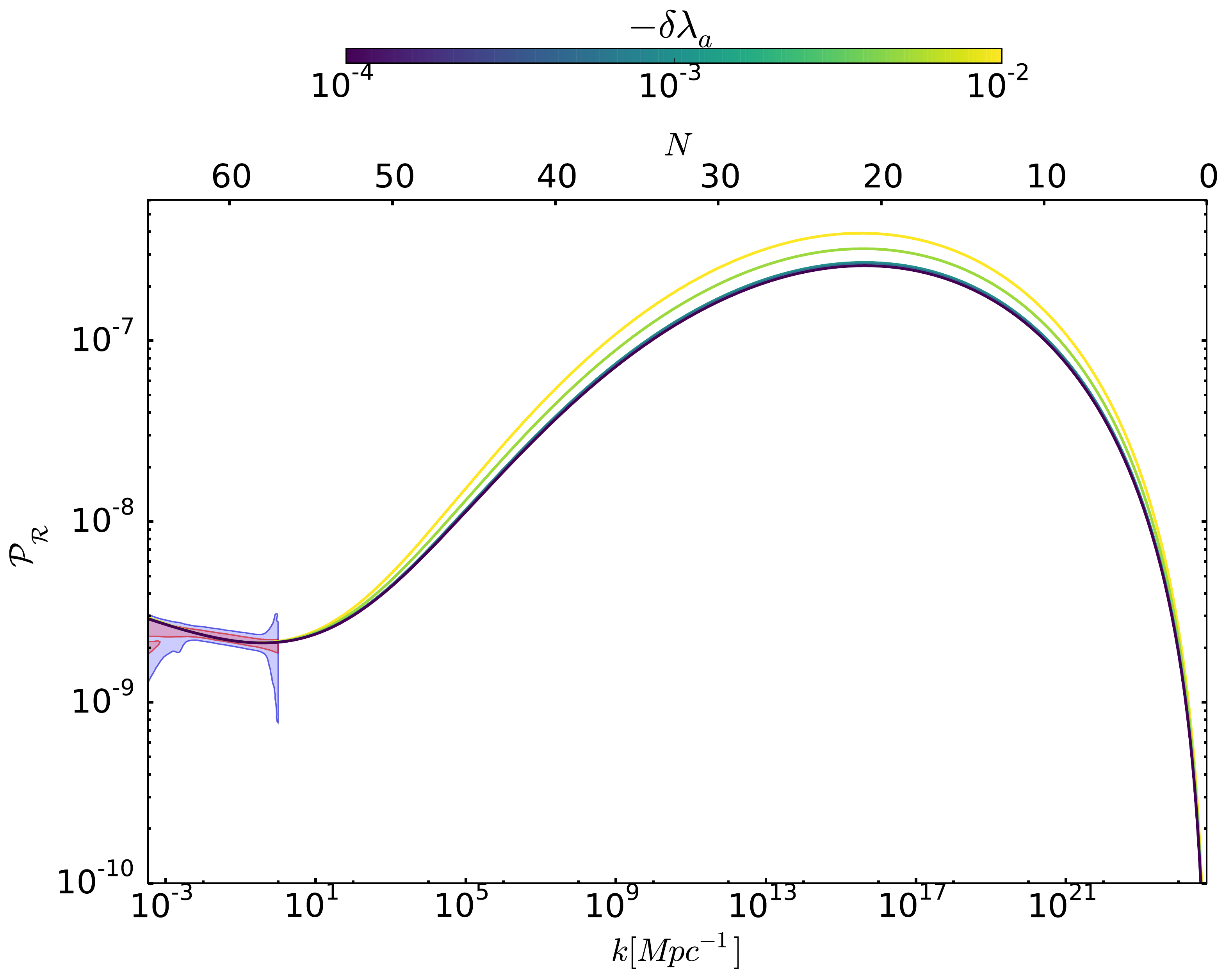}
\caption{The inflationary power spectra ${\cal P}_{\cal R}$ for the \textit{1-loop threshold} scenario as
a function of momenta and the number of e-folds before the end of inflation. Different contours correspond to different values of $\delta\lambda_a$ for the same value of $\xi$.  Here $\kappa$ was varied to fix $n_s=0.968$. The values
of the tensor-to-scalar ratio are again within the Planck 95\% C.L. contour.  The shaded regions at low $k$ mark 
the 68\% and 95\% C.L. constraints on the power spectrum \cite{Ade:2015lrj}.}\label{spectrajumps1loop}
\end{figure}

\begin{figure}
\centering
  \includegraphics[scale=0.248]{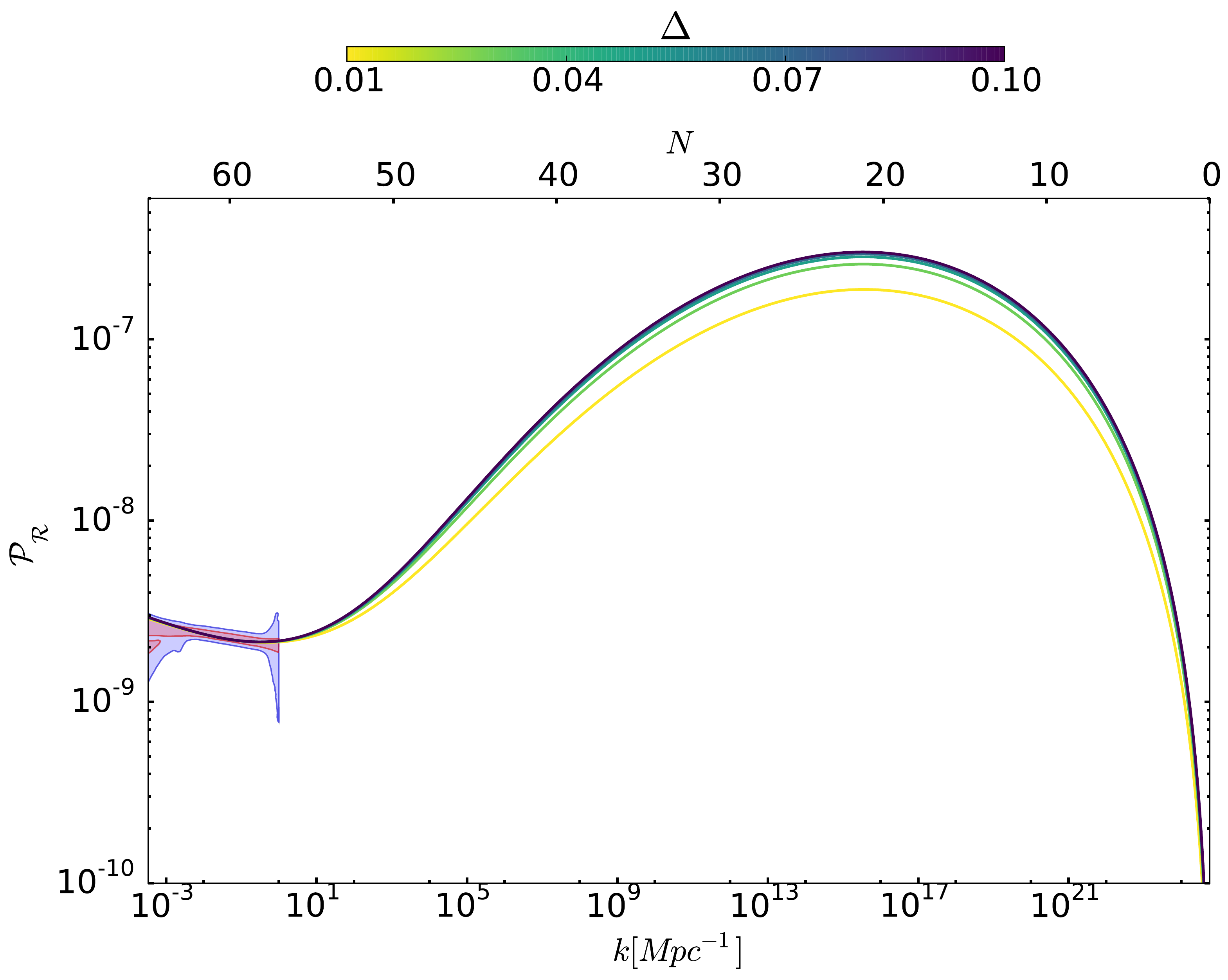}
\caption{The inflationary power spectra ${\cal P}_{\cal R}$ for the \textit{collective threshold} 
scenario with $\delta\lambda =-1 \times 10^{-5}$ as a function of momenta and the number of e-folds 
before the end of inflation. Different contours correspond to different values of $\Delta$ for the
same value of $\xi$.  As before, $\kappa$ was varied to fix $n_s=0.968$.}\label{spectrajumpscollective}
\end{figure}

The non-instantaneous threshold scenarios discussed in Section \ref{sec:spectra2} modify also the shape of the inflationary spectra at small and intermediate scales. As shown in Fig.~\ref{spectrajumps1loop} and \ref{spectrajumpscollective}, these modifications are $\mathcal{O}(1)$ and do not introduce any new phenomenology, such as the creation of primordial black holes after horizon entry \cite{Carr:2016drx, Bird:2016dcv,Carr:2017jsz}. Indeed, the large-amplitude peaks needed for the production of these dark matter candidates are associated with values of the 
tensor-to-scalar ratio that are totally incompatible with observations.
Note however that our analysis explicitly excluded non-monotonic potentials. Considering potentials with a local 
minimum along the inflationary trajectory could generate larger peaks in the spectrum. However, 
the ordinary slow-roll treatment is not appropriate in this situation \cite{Kannike:2017bxn,Germani:2017bcs} and 
further effects, such as the stochastic motion of the inflaton field \cite{Starobinsky:1994bd,Vennin:2015hra,Pattison:2017mbe}, should potentially be considered. We postpone 
the analysis of this issue to a future publication.
  %%%%%%%%%%%%%%%%%%%%%%%%%%%%%
%%%%%%%%%%%%%%%%%%
  %%%%%%%%%%%%%%%%%%%%%%%%%%%%%
%%%%%%%%%%%%%%%%%%

The creation of  a non-negligible amount of primordial black holes in inflationary models with non-monotonic potentials and critical Higgs inflation scenarios  was recently advocated in Refs.~\cite{Garcia-Bellido:2017mdw,Kannike:2017bxn,Ezquiaga:2017fvi}. The analysis presented in this paper displays some 
important differences with Ref.~\cite{Ezquiaga:2017fvi}, in particular: i) the absence of 
threshold effects, ii) the use of a large running for the 
non-minimal coupling $\xi$ which significantly exceeds the usual renormalization group predictions. 
Indeed, the one-loop beta function for $\xi$ in the SM non-minimally coupled to gravity 
is given by \cite{Bezrukov:2009db,Yoon:1996yr}
\begin{equation}
 \beta_\xi(\mu) = \mu \frac{\partial}{\partial \mu} \xi = - \frac{1}{16 \pi^2} \xi \left( \frac{3}{2} g'^2 + 3 g^2 - 6 y_t^2\right) \\.
\end{equation}
For realistic values of the couplings near the Planck scale 
($\xi = 50, g'=0.45, g = 0.5, y_t = 0.4$) the running of $\xi$ is small, $\beta_\xi \propto \mathcal{O}(10^{-2})$. We thus neglected the running in our analysis.
It would be interesting to revisit the claims of  Ref.~\cite{Ezquiaga:2017fvi} in
a well-defined UV completion of the SM non-minimally coupled to gravity, potentially including additional
degrees of freedom able to generate a large $\xi$ running. 
%%%%%%%%%%%%%%%%%%%%%%%%%%%%%
\section{Conclusions}\label{sec:conclusions}
%%%%%%%%%%%%%%%%%%%%%%%%%%%%%
We considered the phenomenological consequences of  higher-dimensional operators respecting the asymptotic symmetries 
of the classical Higgs inflation scenario. Depending on the details of the renormalization group 
running of the Higgs self-coupling up to the inflationary scale one needs to distinguish two cases: 
 \emph{critical Higgs inflation} occurs if the self-coupling \textit{at the inflationary scale} is 
 tiny and the RGE potential develops a  \emph{quasi}-inflection point. 
For larger values of the self-coupling the potential maintains its shape and 
one stays in the regime of \emph{universal Higgs inflation}. Higher-dimensional 
operators modify the SM running by inducing jumps in the coupling
constants within a localized region between the electroweak and the inflationary scales.
Connecting measurements at the electroweak scale to the \textit{critical Higgs inflation} scenario requires
the magnitude of these jumps to be smaller than $\mathcal{O}(10^{-2})$.

Using the RGE potential, we studied the impact of these corrections on the power spectrum of primordial perturbations generated during inflation.  
We considered two particular scenarios:  1) threshold effects consistent with the truncation
of the renormalization group equations at 1-loop and  2) a \textit{purely phenomenological} parametrization that encodes collective threshold effects, smoothly interpolating 
between the low- and high-energy regimes. In spite of their simplicity, these two 
scenarios were enough to illustrate the robustness of \textit{universal Higgs inflation} and the subtleties 
associated with \textit{critical Higgs inflation}.  As long as the RGE potential does not contain 
a \textit{quasi}-inflection point along the inflationary trajectory, the spectrum of primordial 
fluctuations is \textit{universal} and almost insensitive to the particular set of threshold corrections. In 
the presence of an inflection point, the details of the UV completion can no longer be ignored. The 
possible  field-dependence in the vicinity of the \textit{quasi}-inflection point (i.e. at scales close to the 
onset of the scale invariant regime) can modify the shape  of the primordial spectrum. On general 
grounds, precise predictions within \textit{critical Higgs inflation} should be complemented by a particular UV
completion of the Standard Model non-minimally coupled to gravity.

%%%%%%%%%%%%%%%%%%%%%%%%%%%%%
\section*{Acknowledgements}
%%%%%%%%%%%%%%%%%%%%%%%%%%%%%
We thank Georgios Karananas for useful comments on the manuscript. JR 
acknowledges support  from DFG through the project TRR33 ``The Dark Universe''.

\end{document}